\documentclass[sigconf,10pt]{acmart}

\usepackage{enumitem}
\setlist[itemize]{noitemsep,topsep=0pt,wide=\parindent,labelsep=0.5em,align=left}

\usepackage{subcaption}

\usepackage[ruled,vlined,linesnumbered]{algorithm2e}
\SetNlSty{textbf}{\color{blue}}{:}      
\SetCommentSty{itshape\color{gray!70}} 
\newcommand{\pred}[1]{\mathrm{Model}.\textsc{#1}}
\SetKw{Break}{break}        
\SetKw{Continue}{continue}  
\usepackage{xcolor}

\SetCommentSty{myalgcommentstyle}

\usepackage{multirow}
\usepackage{adjustbox}

\AtBeginDocument{%
  }

\setcopyright{acmlicensed}
\copyrightyear{2025}
\acmYear{2025}
\acmDOI{XXXXXXX.XXXXXXX}
\acmConference[Conference acronym 'XX]{Make sure to enter the correct
  conference title from your rights confirmation email}{June 03--05,
  2018}{Woodstock, NY}

\begin{document}

\title{A Predictive and Synergistic Two-Layer Scheduling Framework for LLM Serving}


\author{Yue Zhang}
\affiliation{%
  \institution{Sun Yat-sen University}
  \city{Guangdong}
  \country{China}
}
\email{zhangy2259@mail2.sysu.edu.cn}

\author{Yuansheng Chen}
\affiliation{%
  \institution{Sun Yat-sen University}
  \city{Guangdong}
  \country{China}
}
\email{chenysh253@mail2.sysu.edu.cn}

\author{Xuan Mo}
\affiliation{%
  \institution{Sun Yat-sen University}
  \city{Guangdong}
  \country{China}
}
\email{moxuan@mail2.sysu.edu.cn}

\author{Alex Xi}
\affiliation{%
 \institution{Kaon AI}
 \city{San Francisco}
 \country{USA}
 }
\email{alex@flowgpt.com}

\author{Jialun Li}
\affiliation{%
  \institution{Guangdong Polytechnic Normal University}
  \city{Guangdong}
  \country{China}
}
\email{jialun.li@gpnu.edu.cn}

\author{WeiGang Wu}
\affiliation{%
  \institution{Sun Yat-sen University}
  \city{Guangdong}
  \country{China}
  }
\email{wuweig@mail.sysu.edu.cn}

\renewcommand{\shortauthors}{Zhang et al.}

\begin{abstract}

LLM inference serving typically scales out with a two-tier architecture: a cluster router distributes requests to multiple inference engines, each of which then in turn performs its own internal scheduling. However, this commonly used paradigm suffers from critical, systemic inefficiency caused by the information gaps across two layers. At the cluster-layer, the router mainly relies on lagging, coarse-grained metrics, such as average latency and queue length to make decisions, resulting in ``decision lag'' that leads to suboptimal request routing. At the engine-layer, static heuristic scheduling policies cannot effectively handle the dynamic workloads, leading a poor balance between latency and throughput. Besides, these gaps may cause SLO violations and resource waste, especially in heterogeneous cloud environments.

To bridge such gaps, we propose NexusSched, a cross-layer framework that shifts LLM serving system from reactive load balancing to predictive orchestration. The core of NexusSched lies in a structurally-informed online performance model that provides accurate, forward-looking per-step latency and capacity estimations. This model empowers two key components. At the engine-layer, LENS performs SLO-aware, adaptive scheduling, dynamically optimizing batching to meet SLOs under real-time loads. At the cluster-layer, PRISM uses predictive signals to perform state-driven routing, maximizing cluster-wide performance and SLO attainment. Performance evaluations show that NexusSched improves SLO attainment by 43\% on average and achieves up to 3$\times$ throughput speedup in long-context and heterogeneous scenarios. Besides, we also deploy NexusSched on FlowGPT's clusters to demonstrate its advantages in production environment.

\end{abstract}




\settopmatter{printacmref=false} 
\renewcommand\footnotetextcopyrightpermission[1]{} 
\pagestyle{plain} 
\settopmatter{printccs=false}
\maketitle

\section{Introduction}

Large Language Model (LLM) inference serving has become the critical infrastructure for a wide range of modern AI applications, including chatbots~\cite{adiwardana2020towards,roller2020recipes,ray2023chatgpt}, long-text summarization~\cite{peters2025generalization}, and code generation~\cite{jiang2024survey,zhang2023planning,pujar2023automated}. The autoregressive nature of LLMs, coupled with their immense computational and memory demands, poses significant challenges to achieving low-latency, high-throughput serving~\cite{vaswani2017attention}.

\begin{figure}[!t]
\centering
\includegraphics[width=0.85\columnwidth]{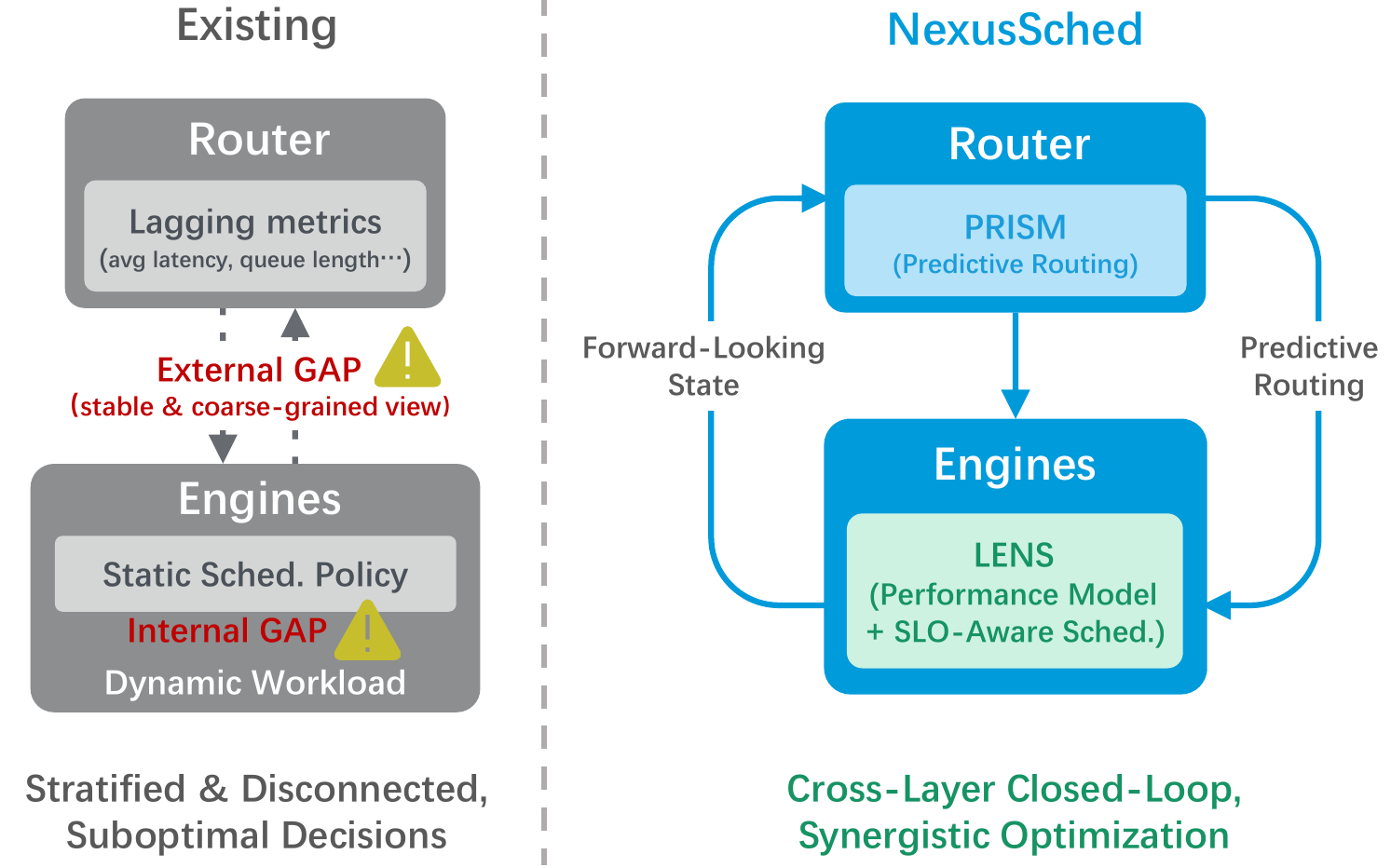}
\caption{Bridging the Information Gaps: Model-Driven Cross-Layer Scheduling.}
\label{fig:intro}
\end{figure}

Existing works mainly focused on vertical scaling via engine-layer optimization. For example, the prefill-priority policy
in vLLM~\cite{kwon2023efficient} can exhibit remarkable effectiveness at minimizing Time-to-First-Token (TTFT) for interactive tasks. Similarly, chunked prefill~\cite{narayanan2024sarathi,holmes2024deepspeed} can successfully improve the overall throughput by interleaving prefill and decode steps. However, one inevitable limitation of these static scheduling policies is the \textbf{internal information gap}. They rely heavily on predefined heuristics (e.g., a static token budget) that cannot accommodate to the dynamic nature of real-world workloads, including shifting request characteristics (e.g., context/generation lengths) and arrival patterns (e.g., burstiness, rate changes). Furthermore, these static policies fail to optimize performance under Service Level Objective (SLO) constraints adaptively. This results in an unavoidable dilemma: either sacrificing Quality of Service (QoS) or resorting to inefficient over-provisioning to handle peak loads.

Apart from vertical scaling, horizontal scaling is also widely adopted to serve massive user requests~\cite{qin2024mooncake,sun2024llumnix} by deploying large clusters with tens or even hundreds of inference instances~\cite{wang2023tabi,oh2024exegpt,wu2023fast}. This paradigm naturally formulates the two-tier scheduling architecture, where an upper-level cluster router acts as a traffic gateway, distributing incoming requests to multiple downstream inference engine instances, and each engine instance performs its own engine-layer batching and scheduling operations~\cite{zhong2024distserve,bin2025fineserve,huang2025slo,goel2025niyamabreakingsilos}.
However, this two-tier architecture introduces one critical \textbf{external information gap} between the router and the engines. Existing routers typically rely on hand-crafted heuristics or coarse-grained, lagging metrics like average latency for load balancing~\cite{zhong2024distserve,bin2025fineserve,huang2025slo,goel2025niyamabreakingsilos}. This creates a significant ``decision lag'', making it difficult for the router to perform reasonable operations, and can result in unpredictable suboptimal placements that cause SLO violations.

The aforementioned two information gaps reveal fundamental, underappreciated systemic inefficiency in the standard LLM serving system. Such inefficiency is further aggravated owing to dynamic workloads and hardware heterogeneity that are commonly adopted in modern cloud environments~\cite{subramanya2023sia,miao2024spotserve}. 
Our core insight is that an accurate, forward-looking prediction mechanism is required to bridge both information gaps, empowering the system to replace reactive, predefined heuristics with proactive, performance-aware orchestration. 

Therefore, we design a \textbf{structurally-informed online performance model}. Existing black-box based approaches, such as polynomial regression models~\cite{hong2025sola,chen2025slosserve}, generalize poorly because that they are agnostic to an instance's service capability, and fail to model the complex, non-linear effects of batching. Instead, our model is structurally-informed, its mathematical structure is inspired by the physical realities of GPU execution, which can accurately model an instance's performance capability across diverse workload deployments. Furthermore, its online learning capability enables a \textbf{``zero-config''} deployment that rapidly adapts to new hardware without costly offline profiling.

Based on the online performance model, we propose \textbf{NexusSched} (shown in Fig.~\ref{fig:intro}), a cross-layer framework that enables predictive decision-making at both layers. At the engine-layer, our performance model enables \textbf{LENS} (Latency-Enhanced Node Scheduler) to serve as a proactive, SLO-aware scheduler. LENS bridges the internal information gap by using the model's latency prediction results to dynamically balance the scheduling performance between latency and throughput. LENS then crafts reliable batching plans that meet SLOs under real-time loads. At the cluster-layer, \textbf{PRISM} (Predictive Routing with Instance State Metrics) uses predictive signals exported from LENS as the forward-looking state vectors to bridge the external gap. PRISM replaces reactive load balancing with predictive request orchestration, using a multi-dimensional scoring function that evaluates signals like predicted latency and pending workload to match each request to the suitable instance.

We systematically evaluated NexusSched on both homogeneous (8$\times$H100) and heterogeneous GPU clusters against state-of-the-art (SOTA) baselines like vLLM~\cite{kwon2023efficient}. Under certain types of real-world scenarios, NexusSched improves SLO attainment by \textbf{43\%} on average and delivers over \textbf{3$\times$} throughput speedup in long-context and heterogeneous scenarios. Furthermore, we also validated NexusSched on production environment from FlowGPT.

In summary, this paper makes the following contributions.
\begin{itemize}
    \item To the best of our knowledge, we are the first to identify and systematically analyze the two fundamental \textbf{information gaps} in LLM serving architectures: an external gap between the cluster router and engines, and an internal gap between static engine policies and dynamic workloads.
    \item We propose a \textbf{structurally-informed online performance model} to bridge both gaps. Its predictive power enables the scheduler to shift from reactive decision-making to proactive, performance-aware orchestration across the serving stack.
    \item We present \textbf{NexusSched}, a cross-layer co-design framework built upon our performance model. It features \textbf{LENS}, an SLO-aware engine scheduler that bridges the internal gap, and \textbf{PRISM}, a predictive router that bridges the external gap, to effectively handle dynamic workloads and hardware heterogeneity.
    \item We implement and evaluate NexusSched to show its advantages over state-of-the-art LLM serving systems.
\end{itemize}

\section{Background and Motivation}

\subsection{Internal Information Gap: Static Scheduling vs. Dynamic Workloads}

The first critical inefficiency lies in the engine itself—an \textbf{information gap between the scheduler's static scheduling policy and the dynamic nature of the workload}. Existing inference engines commonly rely on fixed, heuristic-based strategies to seek for feasible solutions. This is often controlled by a critical parameter, \texttt{num-batched-tokens}, which determines the max number of tokens processed in a single forward pass and thus directly controls the batch size and computational efficiency of each iteration~\cite{kwon2023efficient}.

\begin{figure}[!h]
\centering
\includegraphics[width=0.60\columnwidth]{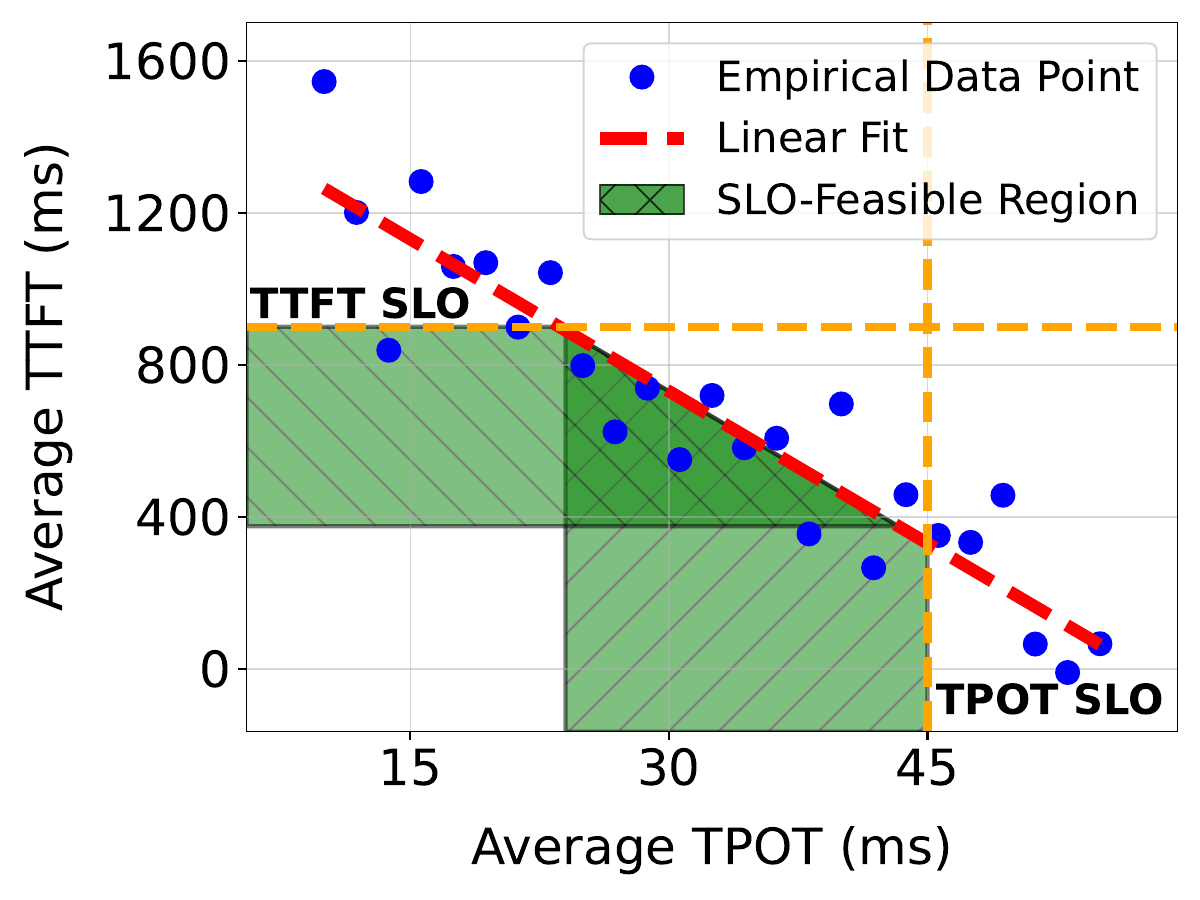}
\caption{Average TTFT and TPOT under different \texttt{num-batched-tokens} settings.}
\label{fig:ttft-tpot-tradeoff}
\end{figure}

To quantify this inherent gap, we measured the average Time-to-First-Token (TTFT) and Time-per-Output-Token (TPOT) under a stable request stream while varying the \texttt{num-batched-tokens} setting. As Fig.~\ref{fig:ttft-tpot-tradeoff} illustrates, the parameter results in a straightforward conflict between these two metrics. A larger \texttt{num-batched-tokens} value allows the system to complete prefill in fewer, more efficient passes, thereby reducing TTFT. However, this is prone to cause head-of-line blocking that worsens the TPOT of concurrent requests. Conversely, a smaller \texttt{num-batched-tokens} value tends to lower TPOT and improve generation throughput, but it significantly increases TTFT by breaking long prompts into multiple inefficient iterations.

A single static configuration cannot satisfy both types of SLOs across the entire load spectrum. This is because that the scheduler is ``uninformed'' about the real-time load and cannot predict the performance of different batch compositions, it is bound to a static rule. This leads to an unnecessary dilemma: either sacrificing QoS for some requests or resorting to inefficient over-provisioning.

\subsection{External Information Gap: Blind Cluster-Layer Routing}
\label{subsec:blindness}

The challenge of local optimization is compounded at the cluster-layer by a fundamental \textbf{external information gap between the router and the individual inference engines}. Production routers are functionally ``blind'', lacking awareness of the real-time, fine-grained state of the engines. They typically rely on coarse-grained, time-aggregated metrics (e.g., 5-second rolling average latency) to make routing decisions.

\begin{figure}[!h]
\centering
\includegraphics[width=0.9\columnwidth]{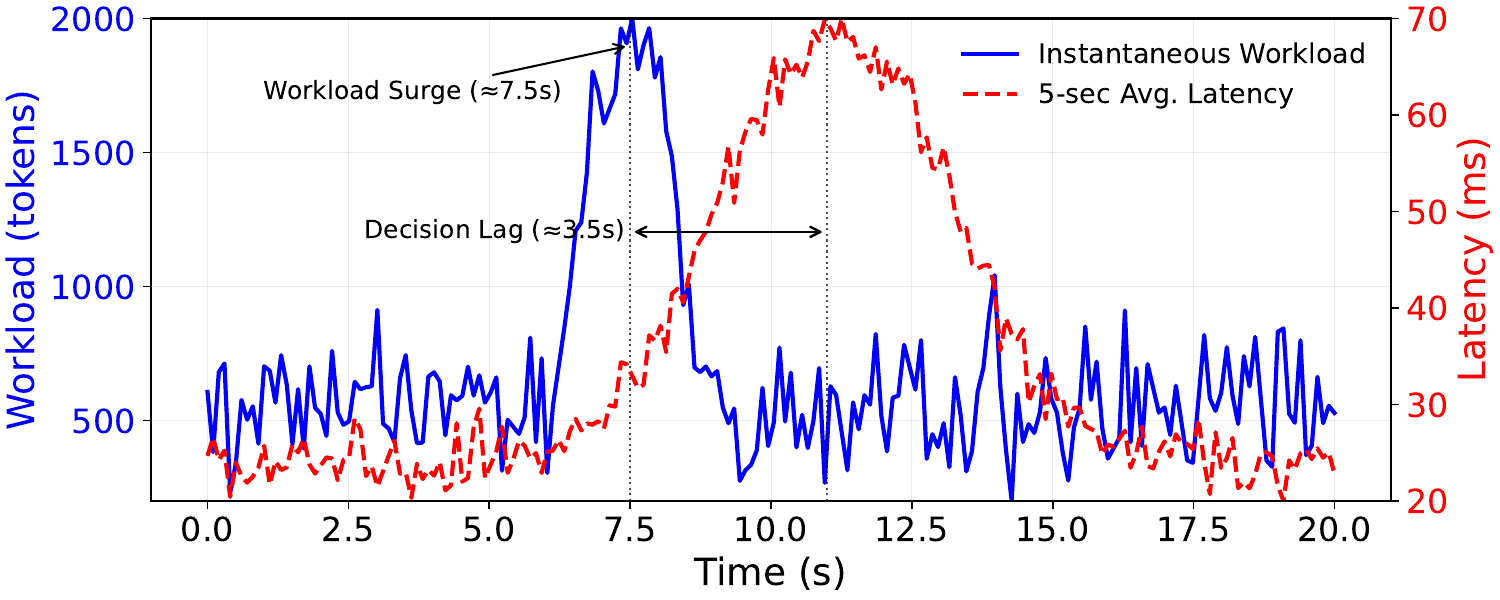}
\caption{Router's Aggregated Latency vs. Engine's Instantaneous Workload.}
\label{fig:router-latency-vs-engine-workload}
\end{figure}

Fig.~\ref{fig:router-latency-vs-engine-workload} quantifies this temporal misalignment, showing how lagging indicators create a \emph{``decision lag''} of several seconds. During this lag, the router with stale information may route latency-sensitive requests to an engine that is already saturated. Such reactive routing fundamentally undermines proactive, capacity-aware load distribution based on past performance and is a primary cause of unpredictable SLO violations.

\subsection{Heterogeneity as an Inefficiency Amplifier}

In modern inference clusters, it has become a common practice to deploy hybrid hardwares (e.g., NVIDIA H100s with A100s/A6000s) to balance infrastructure cost and instance performance. However, such heterogeneity can fundamentally amplify the inherent inefficiency of both engine-layer scheduling and cluster-layer routing.

At the engine-layer, hardware heterogeneity makes static scheduling policies less effective. As shown in Fig.~\ref{fig:runtime-nbt-model-gpu}, the performance curves for different model-hardware combinations diverge significantly. This means that an empirically-tuned heuristic or parameter optimal for one configuration may be severely suboptimal for another. Relying on static, offline profiling to solve every possible combination is operationally infeasible and can incur prohibitive overhead, which highlights the need for dynamic, online adaptation.

\begin{figure}[!t]
\centering
\begin{subfigure}[t]{0.48\columnwidth}
    \centering
    \includegraphics[width=\linewidth]{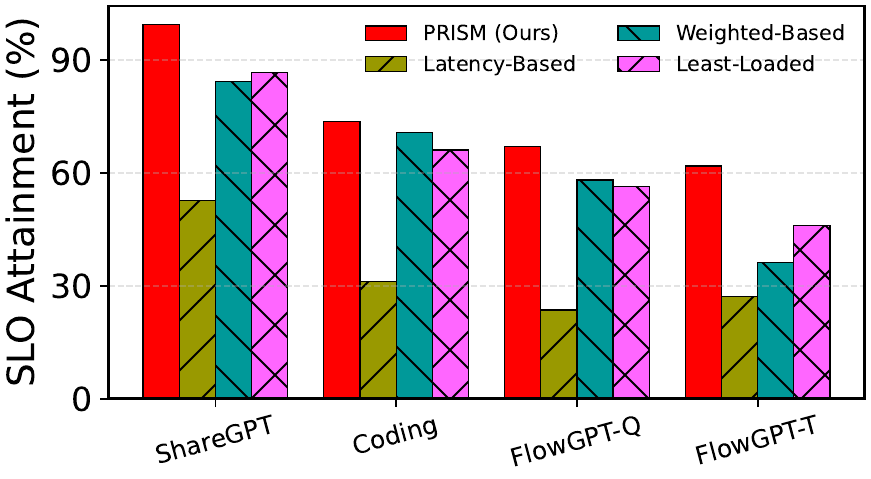}
    \caption{SLO attainment (\%) of different policies on four scenarios under high load. Our policy performs consistently and is superior overall.}
    \label{fig:hete_slo}
\end{subfigure}
\hfill
\begin{subfigure}[t]{0.48\columnwidth}
    \centering
    \includegraphics[width=\linewidth]{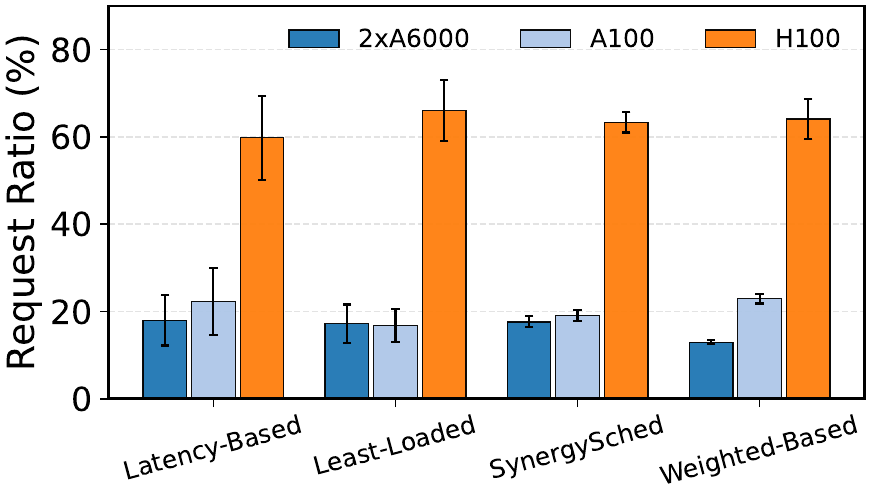}
    \caption{Distribution of request share among heterogeneous instances (H100, 2$\times$A6000, A100) for different scheduling policies in FlowGPT-T scenario.}
    \label{fig:qps_share_flowgpt_ts}
\end{subfigure}
\caption{Overall effectiveness and resource allocation characteristics of different scheduling policies in a heterogeneous environment.}
\label{fig:hete_sched_overview}
\end{figure}

At the cluster-layer, traditional load balancing metrics like queue length can mislead proxy for actual loads when introducing heterogeneity, and greatly degrade the performance.
Even with uniform request lengths, the same queue length implies vastly different processing delays across different GPU types, which inevitably results in poor routing decisions. In contrast, NexusSched's routing uses predictive, forward-looking state to perceive each instance's true service capability. Our experimental results (Fig. ~\ref{fig:hete_sched_overview}) validate this approach, showing PRISM's consistently superior performance in these challenging heterogeneous environments.

\section{System Overview}

\subsection{Architecture and Components}

\begin{figure}[!t]
\centering
\includegraphics[width=1\columnwidth]{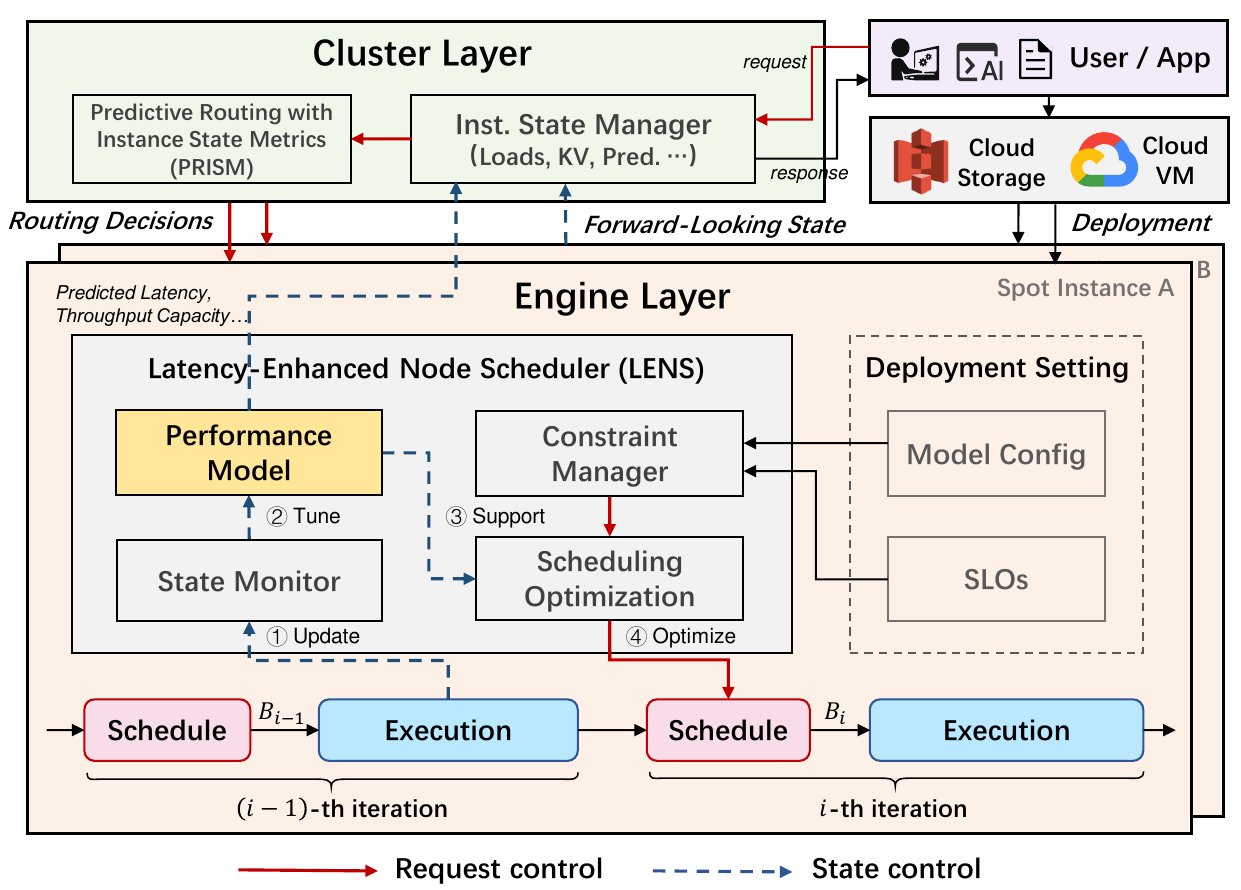}
\caption{NexusSched architecture.}
\label{fig:architecture}
\end{figure}

NexusSched is a cross-layer scheduling framework designed to solve the fundamental information gaps in LLM inference serving. Its core architecture, depicted in Fig.~\ref{fig:architecture}, consists of two synergistic layers: the \textbf{cluster-layer} and the \textbf{engine-layer}. These layers are tightly integrated through the sharing of forward-looking state metrics, which connects two isolated decisions into a single, proactive control loop. Our design decouples macro-level routing from micro-level execution while ensuring that global decisions are grounded in the real-time, predictive reality of each engine.

The \textbf{cluster-layer} serves as the system's entry point, receiving all incoming requests. We present \textbf{PRISM (Predictive Routing with Instance State Metrics)}, a state-driven predictive routing module. It consumes forward-looking state vectors from the engine-layer to select the suitable target engine for each request.

The \textbf{engine-layer} runs on each computing node and is responsible for the actual execution of requests. We present \textbf{LENS (Latency-Enhanced Node Scheduler)}, a proactive intra-node scheduler that bridges the internal information gap. LENS utilizes its internal \textbf{Performance Model} to predict the latency of different batch configurations, selecting the suitable plan under SLO constraints. Crucially, LENS also continuously exports these predictive, forward-looking state metrics to the cluster-layer for global decision-making.

\subsection{End-to-end Request Lifecycle}

The end-to-end process for handling a single request clearly shows how the system's components work together:
\begin{itemize}
    \item \textbf{Arrival and Predictive Routing.} When request $R$ arrives, the cluster-layer \textbf{PRISM} queries for the latest forward-looking state of all engines (e.g., predicted processing latency~$\widehat{L}_e$, pending workload~$W_e$). It then scores candidate engines and dispatches $R$ to the highest-scoring engine $E$ \textit{before} execution begins, enabling proactive routing.

    \item \textbf{Engine Admission and Continuous Optimization.} Upon arrival at $E$, request $R$ joins the local queue and is handled by \textbf{LENS}, which drives a continuous optimization loop. In each scheduling cycle (tens to hundreds of milliseconds), LENS (i) senses queue and memory status via the \textbf{State Monitor}; (ii) sets an adaptive latency target from the SLOs provided by the \textbf{Constraint Manager} and the current load; and (iii) invokes the \textbf{Scheduling Optimizer} to produce the batching plan for the next iteration. This plan specifies which requests (including $R$) will run and how the next iteration is batched (e.g., per-iteration token budget, prefill-slice sizes).

    \item \textbf{Closing the Cross-Layer Loop via State Feedback.} As requests are processed, LENS continuously self-calibrates its performance model with real-time data. More importantly, it exports these refined, forward-looking capacity signals back to PRISM. This feedback closes the control loop, ensuring subsequent cluster-layer routing decisions are based on a precise, near-future prediction of each engine's capacity, not on lagging historical metrics.
\end{itemize}

\section{LENS: Proactive Engine-Layer Scheduling}

In this section, we introduce two core components in LENS to bridge the internal information gap: a structurally-informed performance model (Sec.~\ref{subsec:perf_model}) and an SLO-aware scheduling optimizer (Sec.~\ref{sec:scheduling_algorithm}).

\subsection{Performance Model}
\label{subsec:perf_model}

\begin{figure}[!t]
\centering
\begin{subfigure}[t]{0.48\columnwidth}
    \centering
    \includegraphics[width=\linewidth]{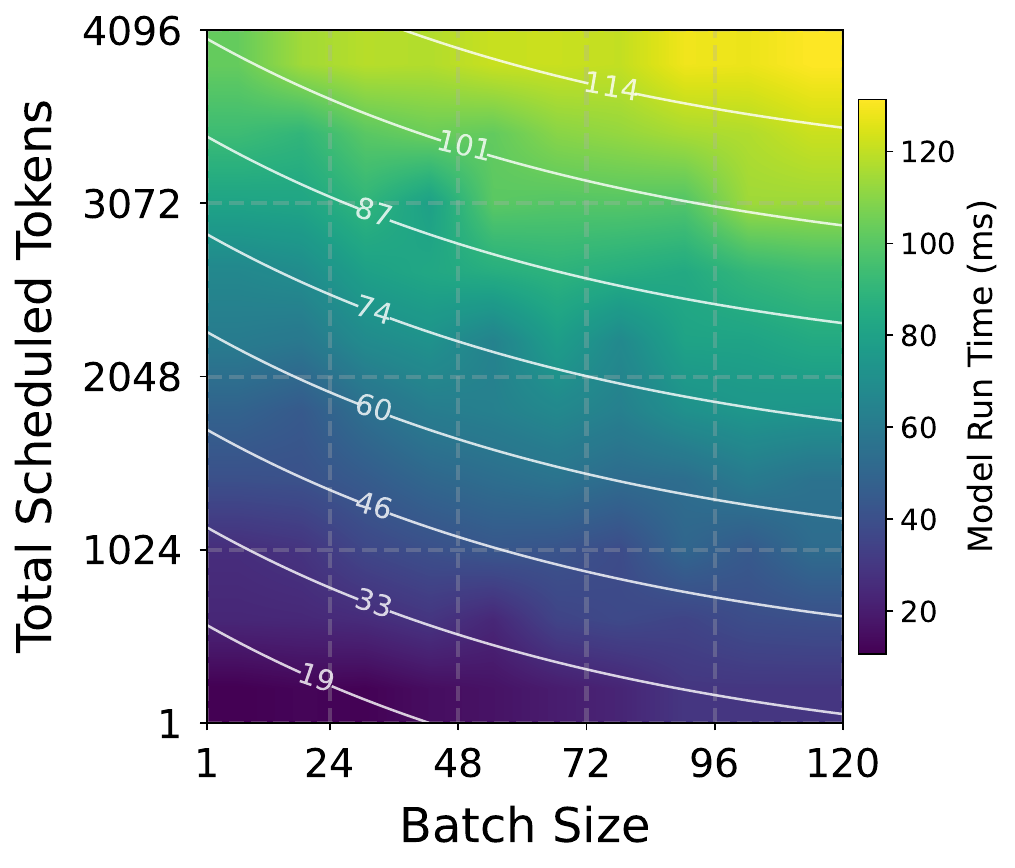}
    \caption{Heatmap of the impact of batch size and total tokens on model run time for Llama3-8B on an H100.}
    \label{fig:batch_token_heatmap}
\end{subfigure}
\hfill
\begin{subfigure}[t]{0.48\columnwidth}
    \centering
    \includegraphics[width=\linewidth]{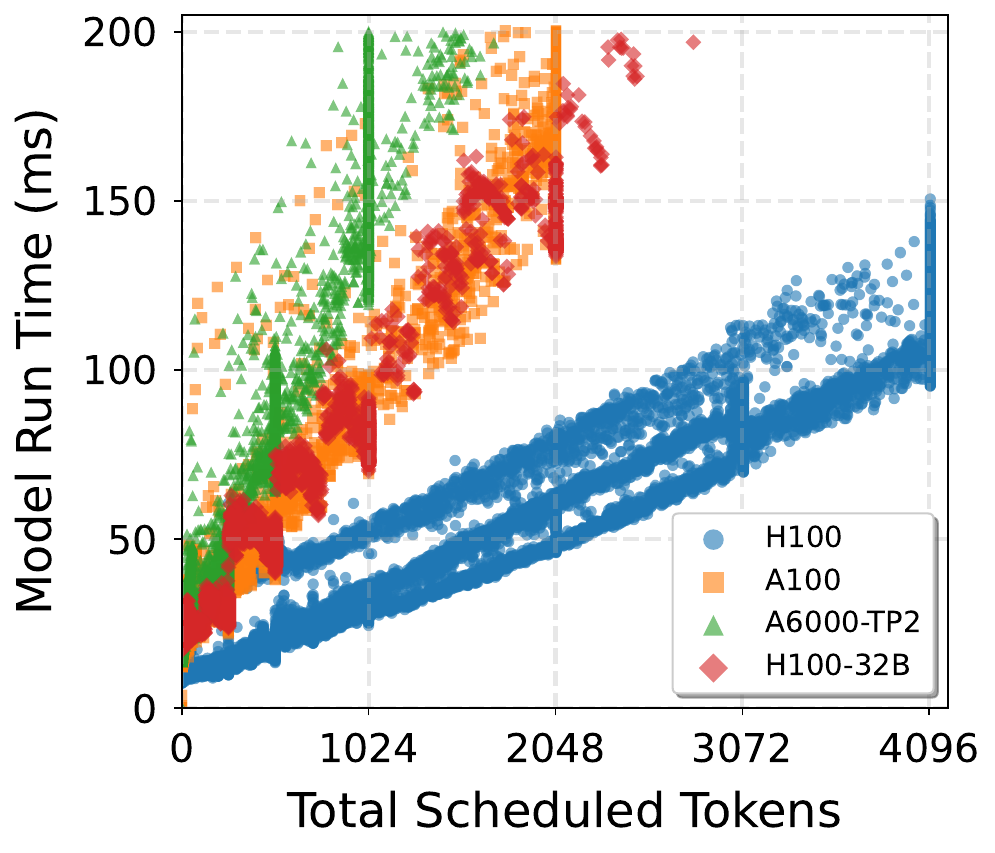}
    \caption{Curve of single-iteration time (Model run time) to total scheduled tokens for different model-GPU combinations (Llama3-8B if unmarked).}
    \label{fig:runtime-nbt-model-gpu}
\end{subfigure}
\caption{Performance analysis of model run time. The two figures together reveal the multi-dimensional factors influencing inference performance.}
\label{fig:profiling}
\end{figure}

Accurate, generalizable performance prediction is the cornerstone of proactive, SLO-aware scheduling. It provides the ``foresight'' that static heuristics lack, which is key to bridging the information gap within the engine. To achieve this, we designed a performance model that not only predicts latency but also perceives an instance's underlying service capability.

We profiled multiple GPUs (H100, A100, A6000) and LLM models (Llama3-8B~\cite{llama3modelcard}, QwQ-32B~\cite{qwq32b}). Fig.~\ref{fig:profiling} highlights two robust behaviors:

\textbf{Workload dependency (token-linear scaling).}
For a fixed batch size $B$, the per-step latency $T$ grows approximately linearly with the total tokens in the batch $S$ (Fig.~\ref{fig:batch_token_heatmap}). This reflects that the total amount of computation is the primary reason to cause latency; $S$ dominates $T$ at fixed $B$.

\textbf{Parallelism saturation (memory-bound at fixed $S$).}
As shown in Fig.~\ref{fig:runtime-nbt-model-gpu}, for a roughly constant total token budget $S$, increasing the batch size $B$ leads to a rise in model run time. The root cause is that higher parallelism triggers more reads/writes and page management for the KV-cache, which lets the memory subsystem (HBM/cache) become the main bottleneck and shifts execution from compute-bound to memory-bound. Meanwhile, as $L=S/B$ decreases, the average prefill sequence becomes shorter. The quadratic FLOPs of attention, $\propto BL^{2}H$ (where $H$ is the LLM model's hidden dimension), converges to $\propto S^{2}H/B$ for a fixed $S$, which indicates that the theoretical compute (FLOPs) will decrease as $B$ increases and exhibits diminishing return.
Once bandwidth approaches saturation, extra parallelism mainly adds memory pressure and scheduling/paging overhead, so runtime still increases but with a diminishing slope (a concave curve). This can be deemed as an upper limit on the GPU's parallel processing and memory subsystem: excessive parallelism pushes throughput into a saturation zone, where the marginal latency penalty decreases as $B$ increases.

These empirical observations directly prove that an effective accurate model should be able to decouple and separately capture the dual effects of total workload (dominated by~$S$) and parallel overhead (influenced by~$B$). Therefore, we build a \textbf{structured model inspired by the physical reality of GPU execution}. Its core idea is to decompose the single-iteration latency~$T(B,S)$~into three interpretable parts: fixed overhead, computation time, and linear overheads related to batch size and sequence length, as shown in Eq.~\ref{eq:latency_overall_rewrite}.

\begin{equation}
T(B,S) = \tau_0 + \frac{\mathrm{Work}(S)}{\mathrm{Thr}(B,S)} + \tau_B B + \tau_S S .
\label{eq:latency_overall_rewrite}
\end{equation}

Here, $\tau_0$~is the load-independent fixed overhead (e.g., kernel launch). The core computation time term is modeled as the quotient of the total workload~$\mathrm{Work}(S)$~and the effective throughput~$\mathrm{Thr}(B,S)$. Based on the linear observation from Fig.~\ref{fig:runtime-nbt-model-gpu}, we approximate the workload as~$\mathrm{Work}(S)=w_0+w_s S$. To capture the saturation phenomenon observed in Fig.~\ref{fig:batch_token_heatmap}, we model the effective throughput~$\mathrm{Thr}(B,S)$~as a dual-exponential saturation function:

\begin{equation}
\mathrm{Thr}(B,S) \;=\; P_{\max}\bigl(1-e^{-k_B B}\bigr)\bigl(1-e^{-k_S S}\bigr).
\label{eq:thr_saturation}
\end{equation}

The structure of Eq.~\ref{eq:thr_saturation} ensures that the modeled throughput, \(Thr(B,S)\), asymptotically approaches the hardware's theoretical peak, \(P_{\max}\), as both parallelism (\(B\)) and total scheduled tokens (\(S\)) increase.
Our model aligns well with residual analysis on multiple platforms and is compatible with other mainstream attention backends besides FlashAttention~\cite{shah2024flashattention3}, such as FlashInfer~\cite{ye2025flashinfer} and XFormers~\cite{xFormers2022}. While simple black-box models (e.g., polynomial fitting) can also predict latency, they lack the instance's capability awareness like \(P_{\max}\), and their predictions are fragile and generalize poorly.

However, a model requiring static, offline calibration would still be fragile in a dynamic cloud environment. Our design overcomes this with a fully online learning and adaptation capability, which enables \textbf{``zero-config''} deployment of NexusSched when facing different LLM models and tasks. 
To balance stability and agility, we draw on the concept of \textbf{multi-timescale adaptation} from control theory~\cite{TUAN1999359}, stratifying the update rates of the model parameters:

\begin{itemize}
    \item \textbf{Structural Parameters ($\{P_{\max}, k_B, k_S\}$)}: These parameters define the basic shape of the performance curve and the physical limits of the hardware, making them relatively stable. We employ a \textbf{low-frequency, long-window} strategy, aggregating data from thousands of batches for non-linear fitting to ensure the model robustly converges to the system's physical characteristics.
    
    \item \textbf{Linear Parameters ($\{w_s, \tau_B, \tau_S\}$)}: These parameters reflect short-term performance drifts caused by the current load, software stack version, etc. We use a \textbf{high-frequency, short-window} strategy, employing lightweight regression on the last few dozen batches for rapid fine-tuning to ensure the model remains sensitive to the current system state.
\end{itemize}

Crucially, all data collection and parameter updates are performed in a separate asynchronous thread, \textbf{imposing no block} on the critical path of serving requests. This online, multi-rate learning framework allows our model to quickly adapt to unseen hardware (like a new spot instance) or models and continuously self-correct to handle performance drift. Experiments (Fig.~\ref{fig:online_learning}) show that the model can converge from an unoptimized state to near-ideal performance within 30 seconds, providing a stable and accurate foundation for the scheduler's decisions.

\subsection{SLO-Aware Scheduling}
\label{sec:scheduling_algorithm}

\begin{table}[t]
\centering
\caption{Summary of notation used in the LENS.}
\label{tab:lens_notation_detailed}
{\small
\begin{tabular}{l|l|l}
\hline
\textbf{Class} & \textbf{Symbol} & \textbf{Explanation} \\
\hline
\multirow{5}{*}{Setting} & $\mathrm{SLOs}$ & Service Level Objectives for requests \\
 & $\tau_p^{\mathrm{SLO}}$ & The SLO constraint for TTFT \\
 & $\tau_d^{\mathrm{SLO}}$ & The SLO constraint for TPOT \\
 & $M_{\max}$ & Maximum token per batch \\
 & $Q_{\max}$ & Maximum requests per batch \\
\hline
\multirow{2}{*}{System State} & $Q_i^{\mathrm{wait}}$ & Queue of waiting queries at step $i$ \\
 & $Q_i^{\mathrm{run}}$ & Queue of running queries at step $i$ \\
\hline
\multirow{5}{*}{Metrics} & $t_p$ & Avg. TTFT in ms \\
 & $t_d$ & Avg. TPOT in ms \\
 & $\bar{L}$ & Expected decode length in tokens \\
 & $t_d^{\mathrm{TTFT}}$ & Target TPOT under TTFT SLO \\
 & $t_d^{\mathrm{TPOT}}$ & Target TPOT under TPOT SLO \\
\hline
Component & $\mathrm{Model}$ & The performance prediction model \\
\hline
\end{tabular}
}
\end{table}

\begin{algorithm}[t]
\small
\caption{SLO-Aware Scheduling in LENS}
\label{alg:SLO-aware-plus}

\KwIn{%
    $Q_i^{\mathrm{wait}}$,
    $Q_i^{\mathrm{run}}$,
    $\mathrm{SLOs}$,
    $M_{\max}$,
    $Q_{\max}$,
    $\mathrm{Model}$
}
\tcp{Hyperparameters: $N_{search\_iters}$, $\varepsilon_{ratio}$}
\KwOut{best token allocation $\mathcal{A}^{\star}$}

\BlankLine
\If{$Q_i^{\mathrm{wait}} = \emptyset \land Q_i^{\mathrm{run}} = \emptyset$}{
    \Return $\emptyset$\;
}

$T_{\mathrm{adaptive}} \gets
 \textsc{TargetLatency}\bigl(Q_i^{\mathrm{wait}}\cup Q_i^{\mathrm{run}},\;
 \mathrm{SLOs}\bigr)$\;

$\mathcal{A}^{\star} \gets \emptyset$\;
$\mathrm{min\_error} \gets \infty$\;

\For{$B = |Q_i^{\mathrm{run}}|$ \KwTo
     $\min\!\bigl(|Q_i^{\mathrm{run}}|+|Q_i^{\mathrm{wait}}|,\; Q_{\max}\bigr)$}{

    \tcp{Binary search for suitable token budget S}
    $S_{low} \gets B$;\quad $S_{high} \gets M_{\max}$;\quad $S \gets B$\;
    \For{$iteration = 1$ \KwTo $N_{search\_iters}$}{
        \If{$S_{low} > S_{high}$}{\Break\;}
        $S_{mid} \gets \lfloor(S_{low} + S_{high}) / 2\rfloor$\;
        $T_{pred} \gets \pred{Predict}(B, S_{mid})$\;
        \If{$T_{pred} \le T_{\mathrm{adaptive}}$}{
            $S \gets S_{mid}$;\quad $S_{low} \gets S_{mid} + 1$\;
        }
        \Else{
            $S_{high} \gets S_{mid} - 1$\;
        }
    }
  
    \tcp{Greedy allocation based on found budget S}
    $\mathcal{A} \gets \textsc{AllocateTokens}\bigl(Q_i^{\mathrm{run}},\;
     Q_i^{\mathrm{wait}},\; B,\; S\bigr)$\;

    $T_{\mathrm{pred}}^{\star} \gets \pred{Predict}(\mathcal{A})$;
    
    $\mathrm{error} \gets |T_{\mathrm{pred}}^{\star} - T_{\mathrm{adaptive}}|$\;

    \If{$\mathrm{error} < \mathrm{min\_error}$}{
        $\mathrm{min\_error} \gets \mathrm{error}$\;
        $\mathcal{A}^{\star} \gets \mathcal{A}$\;
        \If{$\mathrm{min\_error} < T_{\mathrm{adaptive}} \times \varepsilon_{ratio}$}{
            \Break\;
        }
    }
}
\end{algorithm}

The Scheduling Optimizer is the decision-maker core of LENS. Tab. ~\ref{tab:lens_notation_detailed} summarizes the notations used in this section. It formulates decisions using a dynamic optimization algorithm (see Algorithm~\ref{alg:SLO-aware-plus}) that is both \textbf{SLO-aware} and \textbf{load-adaptive}. This algorithm translates user-defined SLOs into a performance budget that dynamically adjusts to real-time load and solves a constrained optimization problem to find the suitable batching plan that maximizes throughput. This shifts the engine scheduler from passive execution to proactive scheduling.

\subsubsection{Scheduling Algorithm}
The algorithm is invoked at the start of each iteration, prior to model execution.
First, it obtains a real-time snapshot of the engine's internal state—the current running queue $Q_i^{\mathrm{run}}$ and waiting queue $Q_i^{\mathrm{wait}}$—via the State Monitor component. Next, it parses the SLO objects from the Constraint Manager and, together with the current load, calls \textsc{TargetLatency} (Sec.~\ref{subsubsec:compute-target-latency-cn}) to produce the adaptive latency target \(T_{\mathrm{adaptive}}\) (line~3).

The core of the algorithm is a constrained search process (lines 6--24) aimed at finding the suitable batch configuration for the next execution step. The process iterates through all reasonable batch sizes $B$ as candidate solutions. For each candidate $B$, instead of performing a complex inverse calculation, the scheduler leverages the monotonicity of the performance model to efficiently find the suitable token budget $S$ that can be processed within the adaptive target latency $T_{\mathrm{adaptive}}$ via a binary search (lines 8--16). 

After determining this target configuration $(B, S)$, the algorithm constructs the actual candidate batch $\mathcal{A}$ using a greedy policy (line~17). This policy adheres to the principle of continuous batching by prioritizing requests already in progress, and then adding new requests from the waiting queue in a FCFS order until the limits of either $B$ or $S$ are reached. Once constructed, the algorithm again uses the model's forward prediction capability to accurately calculate the expected latency of this actual batch, $T_{\mathrm{pred}}^{\star}$ (line~18), and evaluates its error relative to the target latency (line~19). Finally, the algorithm selects the batch configuration among all candidate batches that most closely matches the dynamic target as its final decision.

While the theoretical time complexity is $O(N_{wait} \cdot Q_{max})$, the practical overhead is negligible (Fig.~\ref{fig:scheduling_overhead}). This efficiency stems from the fact that both the number of waiting requests, $N_{wait}$, and the maximum batch size, $Q_{max}$, are small and bounded in a limited serving engine, while the internal binary search is constrained to a maximum of 10 iterations. This efficiency is further improved by an early-exit mechanism (lines 23--24) that terminates the search once a sufficiently suitable solution is found.

\subsubsection{SLO-Aware Target Selection}
\label{subsubsec:compute-target-latency-cn}

\noindent Given the SLO constraints on TTFT and TPOT, we select a \emph{target TPOT} \(t_d\) (ms) for the next iteration that lies within the feasible region and strikes an appropriate balance between responsiveness (TTFT) and sustained throughput (TPOT) for the current workload. Let \(t_p\) and \(t_d\) denote the average TTFT and TPOT values. This choice directly determines the batch shape and size of the subsequent iteration, and is therefore central to \textsc{SLO-Aware Scheduling}.

\textbf{A modelable TTFT–TPOT relationship.}
Under continuous batching with chunked prefill, the measured TTFT \((t_p)\) decreases approximately linearly with increasing TPOT \((t_d)\) within the operating range (Fig.~\ref{fig:ttft-tpot-tradeoff}). We model this relationship with a function \(\hat{t}_p(t_d)\), which estimates the TTFT for a given TPOT:
\begin{equation}
  \hat{t}_p(t_d) \approx \alpha - \beta\, t_d, \qquad \beta>0 .
\end{equation}

Here, \(\alpha\) is the theoretical maximum TTFT (the y-intercept), and \(\beta\) is the trade-off rate between TTFT and TPOT.

For a request with expected decode length \(\bar L\) (tokens), the expected end-to-end latency is
\begin{equation}
  T_{\mathrm{e2e}}(t_d) \;=\; \hat{t}_p(t_d) + \bar L \cdot t_d
  \;=\; \alpha + (\bar L-\beta)\, t_d .
\end{equation}

\textbf{SLO-feasible region and boundaries.}
Let \(\tau_p^{\mathrm{SLO}}\) denote the TTFT SLO and \(\tau_d^{\mathrm{SLO}}\) the TPOT SLO. Feasibility requires
\begin{equation}
  \hat{t}_p(t_d) \le \tau_p^{\mathrm{SLO}}
  \ \Rightarrow\
  t_d \ge t_d^{\mathrm{TTFT}} \triangleq \frac{\alpha - \tau_p^{\mathrm{SLO}}}{\beta},
\end{equation}
\begin{equation}
  t_d \le t_d^{\mathrm{TPOT}} \triangleq \tau_d^{\mathrm{SLO}} .
\end{equation}

Hence, the target \(t_d\) must satisfy \(t_d \in [\,t_d^{\mathrm{TTFT}},\, t_d^{\mathrm{TPOT}}\,]\) (see Tab.~\ref{tab:lens_notation_detailed}). If the interval is empty, we fall back to the nearest boundary and report a potential SLO risk.

\textbf{Workload-aware decision rule and adaptive heuristic.}
For a given workload, where \(\bar L\) is estimated from serving history, minimizing \(T_{\mathrm{e2e}}\) under these constraints yields a simple, implementable policy. The decision hinges on the relationship between the expected decode length (\(\bar L\)) and the trade-off rate (\(\beta\)): when \(\bar L > \beta\) (long outputs, TTFT pressure dominates), end-to-end latency is minimized by choosing the smallest feasible \(t_d\). Therefore, we set the target at the TTFT boundary, \(t_d^{\mathrm{TTFT}}\). When \(\bar L \le \beta\) (short outputs, where increasing \(t_d\) significantly reduces TTFT), we choose the largest feasible \(t_d\), which is the TPOT boundary, \(t_d^{\mathrm{TPOT}}\).
This logic is formalized in Eq.~\ref{eq:adaptive}, where \(t_{d,\min}\) is a minimum safety value set in practice:
\begin{equation}
  T_{\mathrm{adaptive}}=
  \begin{cases}
    \min\!\bigl(t_d^{\mathrm{TPOT}},\ \max(t_{d,\min},\,t_d^{\mathrm{TTFT}})\bigr), & \bar L>\beta,\\[4pt]
    t_d^{\mathrm{TPOT}}, & \bar L\le\beta.
  \end{cases}
\label{eq:adaptive}
\end{equation}

To accommodate online fluctuations, when the load rises (e.g., the local waiting queue is non-empty), we \emph{gradually relax} the target by \emph{smoothly moving} \(T_{\mathrm{adaptive}}\) upward toward \(t_d^{\mathrm{TPOT}}\) within the feasible region. This prioritizes TTFT and avoids extra latency caused by queuing.


\section{Cluster-Layer Routing Design}
\label{sec:router}

In this section, we introduce two core components to bridge the external information gap: a set of \textbf{forward-looking state metrics} that provide predictive insights (Sec.~\ref{sec:state_sharing_metrics}) and the \textbf{PRISM routing algorithm} that consumes these metrics to make globally optimal decisions (Sec.~\ref{sec:prism_algorithm}).

\subsection{State Sharing for Predictive Routing}
\label{sec:state_sharing_metrics}

To enable predictive routing, our design is founded on a set of \textbf{forward-looking state metrics} exported by each LENS-enabled engine. These metrics serve as an ``information bridge'', providing the cluster-layer with a precise, near-future view of each engine's real capacity and availability. This allows the router to make proactive decisions based on what an engine will be able to handle, a more reliable foundation for routing than traditional lagging indicators like past average latency or queue length.

At each moment~$t$, each engine~$e$~exposes a state vector~$\mathbf{s}_e(t)$~to the routing layer:
\begin{equation}
\mathbf{s}_e(t) = \left\langle \widehat{L}_e, W_e, M_e^{\text{free}}, P_{e, \max} \right\rangle .
\label{eq:clss_vector}
\end{equation}

Each metric in this vector is carefully chosen to empower a key decision-making capability in PRISM:

\begin{itemize}
    \item $\widehat{L}_e$: \textbf{Predicted Processing Latency}. This is the most important forward-looking signal: the model's prediction for the execution time of the \textit{current} batch. It directly quantifies the engine's immediate computational load and the queuing delay a new request would face.
    
    \item $W_e$: \textbf{Pending Workload}. A composite metric capturing both future computational pressure (total pending prefill tokens) and concurrency pressure (number of requests in running and waiting queue).
    
    \item $M_e^{\text{free}}$: \textbf{Effective Available KV Cache}. The available memory \textit{after} accounting for the estimated requirements of already queued requests, providing a reliable signal for admission control.
    
    \item $P_{e, \max}$: \textbf{Peak Throughput Capacity}. A static scalar from the performance model representing the node's theoretical performance limit, which is essential for normalizing load across heterogeneous hardware.
\end{itemize}

\subsection{PRISM: Predictive Routing}
\label{sec:prism_algorithm}

Armed with the \textbf{real-time, forward-looking state exported by each engine}, PRISM solves a single-objective problem: \emph{maximize throughput subject to SLO constraints}. To solve this efficiently on the latency-critical path, PRISM employs a carefully designed scoring function, $\text{score}(e,r)$, which transforms the complex optimization problem into an evaluation of each candidate engine $e$. This scoring function serves as a multi-dimensional decision model designed to match each incoming request $r$ with the most suitable engine:

\begin{equation}
\text{score}(e,r) = \prod_{i=1}^{4} S_i(e,r)^{w_i} .
\label{eq:PRISM_score}s
\end{equation}

We chose a multiplicative form over an additive one because scheduling decisions often exhibit a ``bottleneck effect'': an extreme disadvantage in any single dimension (e.g., severe latency overshoot) should cause the engine's final score to drop sharply, leading to its rejection. The weights $w_i$ are used to adjust the relative importance of different scheduling objectives. In our implementation, all weights default to 1, giving each dimension equal importance. However, the framework supports empirical tuning of these weights; for example, in scenarios dominated by long sessions, the weight corresponding to $S_{\text{affinity}}$ could be increased to prioritize the performance gains from KV cache reuse.



The scoring function has four dimensions, each designed to evaluate a specific objective using the forward-looking state metrics:


\begin{itemize}
    \item \textbf{SLO Margin Assessment ($S_{\text{latency}}$)}: This term evaluates the engine's immediate responsiveness. It uses the predicted processing latency~$\widehat{L}_e$—the predicted time for the engine to complete its current batch—to measure the ``enqueueing delay'' a new request will face. The function compares this latency to the request's SLO; a lower~$\widehat{L}_e$~results in a high score close to 1, while a higher value causes the score to plummet. This directly links the engine's immediate congestion to the QoS target.

    
    \item \textbf{Relative Load Assessment ($S_{\text{load}}$)}: This term assesses the engine's medium-to-long-term load pressure. It uses our composite metric~$W_e$, which captures both future computational pressure and current concurrency pressure. After normalizing by the peak throughput capacity~$P_{e, \max}$~($\rho_e = W_e / P_{e, \max}$), this scoring component penalizes engines with higher relative loads, ensuring new requests are directed to nodes with \emph{relatively higher} capacity.

    
    \item \textbf{Capacity Admission Assessment ($S_{\text{capacity}}$)}: This term acts as a reliability gatekeeper to ensure scheduling success. It uses the effective available KV cache~$M_e^{\text{free}}$—the value after accounting for capacity reserved for queued requests—to determine if the engine can accommodate a new request. The score approaches 1 only when~$M_e^{\text{free}}$~is significantly larger than the request's estimated demand; otherwise, it trends towards 0, fundamentally preventing performance degradation from request preemption.

    
    \item \textbf{Session Affinity Assessment ($S_{\text{affinity}}$)}: This term is designed to optimize performance for conversational applications. By default, its value is a neutral 1. If an engine~$e$~has already cached the historical KV cache for request~$r$, its score is boosted to a constant~$\beta > 1$, incentivizing the router to select nodes where cache reuse is possible.

\end{itemize}

Finally, PRISM selects the engine with the highest score to serve the request. Through this multi-dimensional, state-driven, and objective-oriented scoring mechanism, PRISM achieves fine-grained, predictive scheduling of cluster-wide resources at the request level.




\section{Implementation}

We implemented NexusSched as a modular extension to the vLLM serving ecosystem, using Python and deployed via Kubernetes. Our implementation is divided into two parts: an enhanced engine layer (LENS) and a standalone cluster router (PRISM).

\noindent \textbf{LENS Implementation.} We extended vLLM with approximately 4,000 lines of code to implement LENS. Key modifications include: a \textbf{pluggable scheduler design} to replace vLLM's native heuristic with our SLO-aware algorithm; an asynchronous, non-blocking \textbf{online learning mechanism} for the performance model; and a lightweight, UDP-based \textbf{asynchronous state reporting} module to export predictive metrics to the router with negligible overhead. 

\noindent \textbf{Cluster-Router-Layer Implementation.} The router is a standalone microservice built on vLLM's Production-Stack for compatibility with standard cloud deployments. With approximately 1,500 lines of code, we implemented the PRISM routing algorithm as a \textbf{pluggable policy}, including graceful degradation logic to handle stale state from engines. This decoupled design allows our router to manage containerized engines that export the required forward-looking state metrics.

Our modular designs ensure broad compatibility with mainstream serving ecosystems. For instance, LENS can be adapted to other continuous batching engines like SGLang~\cite{zheng2024sglang} and TensorRT-LLM~\cite{vaidya2023tensorrt}, while the decoupled router can manage a diverse range of engine that exports the required forward-looking state metrics.

\section{Evaluation}

In this section, we present a comprehensive experimental evaluation of NexusSched to validate its effectiveness and robustness. Our experiments are designed to answer the following key questions:

\begin{itemize}
    \item How does NexusSched's end-to-end performance compare to SOTA systems under various workloads (Sec.~\ref{sec:e2e})?
    \item What are the individual contributions of NexusSched's LENS and PRISM? Is the cross-layer co-design truly indispensable for achieving optimal performance (Sec.~\ref{sec:ablation})?
    \item How efficiently does NexusSched manage resources while sustaining high performance in heterogeneous clusters (Sec.~\ref{sec:heterogeneous_cluster_performance})?
    \item When faced with a new, unprofiled GPU or LLM model, can NexusSched's online learning capabilities enable it to quickly converge to ideal performance (Sec.~\ref{sec: zero-config})?
    \item How accurate is our performance model, and what is the real-world overhead introduced by our scheduling components (Sec.~\ref{sec: micro})?
\end{itemize}


\subsection{Experimental Setup}
\label{sec:setup}

\noindent \textbf{Testbed.}
We define a cluster as a set of inference instances managed by a single router. Our experiments were conducted in both homogeneous and heterogeneous cluster to comprehensively evaluate our system's performance.
\begin{itemize}
    \item \textbf{Homogeneous Cluster}: Used for end-to-end performance and ablation studies, consisting of eight NVIDIA H100 (80GB) GPUs connected via NVLink.
    \item \textbf{Heterogeneous Cluster}: Used to test capability-aware routing, composed of a mix of one NVIDIA H100 (80GB), one NVIDIA A100 (80GB), and two NVIDIA A6000 (48GB) GPUs, simulating a common hardware configuration in cloud environments.
\end{itemize}

We evaluated two representative models widely used in both industry and academia: Llama3-8B and QwQ-32B (FP8-quantized). Unless stated otherwise, all experiments ran in a single-model-per-GPU replicated configuration, with the sole exception of Llama3-8B, which used 2-way tensor parallelism across two A6000 GPUs.


\noindent \textbf{Workloads.}
To ensure the realism and comprehensiveness of our evaluation, we adopted a strategy that combines production workloads with diverse public workloads.
\begin{itemize}
    \item \textbf{Production Workloads (FlowGPT Trace)}: This is the core of our evaluation. We collected partial user traffic from FlowGPT's real production system. The value of this trace lies in its characteristics of real-world LLM chatbot services: the burstiness of request arrivals and authentic user request patterns. We replayed this trace in two modes: timestamp(T) mode strictly adheres to the original arrival patterns to test the system's ability to handle bursty loads, while qps(Q) mode sends requests at a uniform rate to evaluate the system's steady-state performance.
    \item \textbf{Public Workloads}: We used three widely recognized public workloads: arXiv-Summarization~\cite{huggingface2025arxivsum} (long-text summarization), Code-Reasoning~\cite{huggingface2025codereasoning} (code generation), and ShareGPT~\cite{huggingface2025sharegpt} (dialogue). As shown in Tab.~\ref{tab:workload_stats}, these workloads cover a range of request characteristics from prefill-heavy to decode-heavy, thoroughly testing the system's performance in different scenarios.
\end{itemize}


\begin{table}
\centering
\caption{Statistics of Workloads.}
\label{tab:workload_stats}
\small
\begin{adjustbox}{width=\columnwidth}
\begin{tabular}{l|ccc|ccc}
\hline
         & \multicolumn{3}{c|}{\textbf{Prompt Tokens}} & \multicolumn{3}{c}{\textbf{Output Tokens}} \\
Workload  & Mean & P99 & Std.                           & Mean & P99 & Std. \\ \hline
FlowGPT  & 4089 & 7014 & 2061                          & 177  & 51  & 200  \\
Coding    & 440  & 1131 & 214                           & 283  & 1096& 233  \\
ShareGPT & 370  & 1420 & 351                           & 249  & 760 & 170  \\
Summarization    & 8936 & 10171& 694                           & 259  & 587 & 115  \\ \hline
\end{tabular}
\end{adjustbox}
\end{table}

\noindent \textbf{Baselines.}
We compare NexusSched against SOTA production solutions. For our multi-dimensional comparison, we select representative benchmarks at both the engine and router layers and combine them to form powerful end-to-end baseline systems.
\begin{itemize}
    \item \textbf{Engine-Layer Baselines}: (i) \textbf{vLLM (v0.9.2, v0)}, a widely adopted industry baseline whose default scheduling policy prioritizes prefill requests to optimize TTFT. (ii) \textbf{Sarathi-Serve}, proposed chunked prefill and later integrated into vLLM. We use the vLLM (v0.9.2, v1) as the representative baseline for chunked prefill to optimize throughput.
    \item \textbf{Router-Layer Baselines}: (i) \textbf{Round-Robin (RR)}, the most common load balancing policy. (ii) \textbf{Session Affinity}, which schedules based on session ID, a mainstream strategy for optimizing KV cache hit rates in conversational scenarios.
\end{itemize}
In our experiments, we compare NexusSched against the combinations of these components, such as \textbf{vLLM RR} (a general-purpose baseline) and \textbf{Sarathi Session} (a SOTA baseline for conversational scenarios).


\noindent \textbf{Metrics.}
We focus on P50, P90 end-to-end latency, P50 TTFT, P50 TPOT, and the SLO attainment. The SLO attainment is defined as the percentage of requests successfully completed within the SLOs and is the core metric for measuring the quality of service under a specific load. Following production practices, we set specific SLOs for different scenarios. For example, for the interactive FlowGPT scenario, we set a TPOT SLO of 12\,ms; for the compute-intensive summarization scenario, the TTFT SLO is 2\,s.


\subsection{End-to-End Performance}
\label{sec:e2e}

\begin{figure*}[!t]
\centering
\includegraphics[width=1.9\columnwidth]{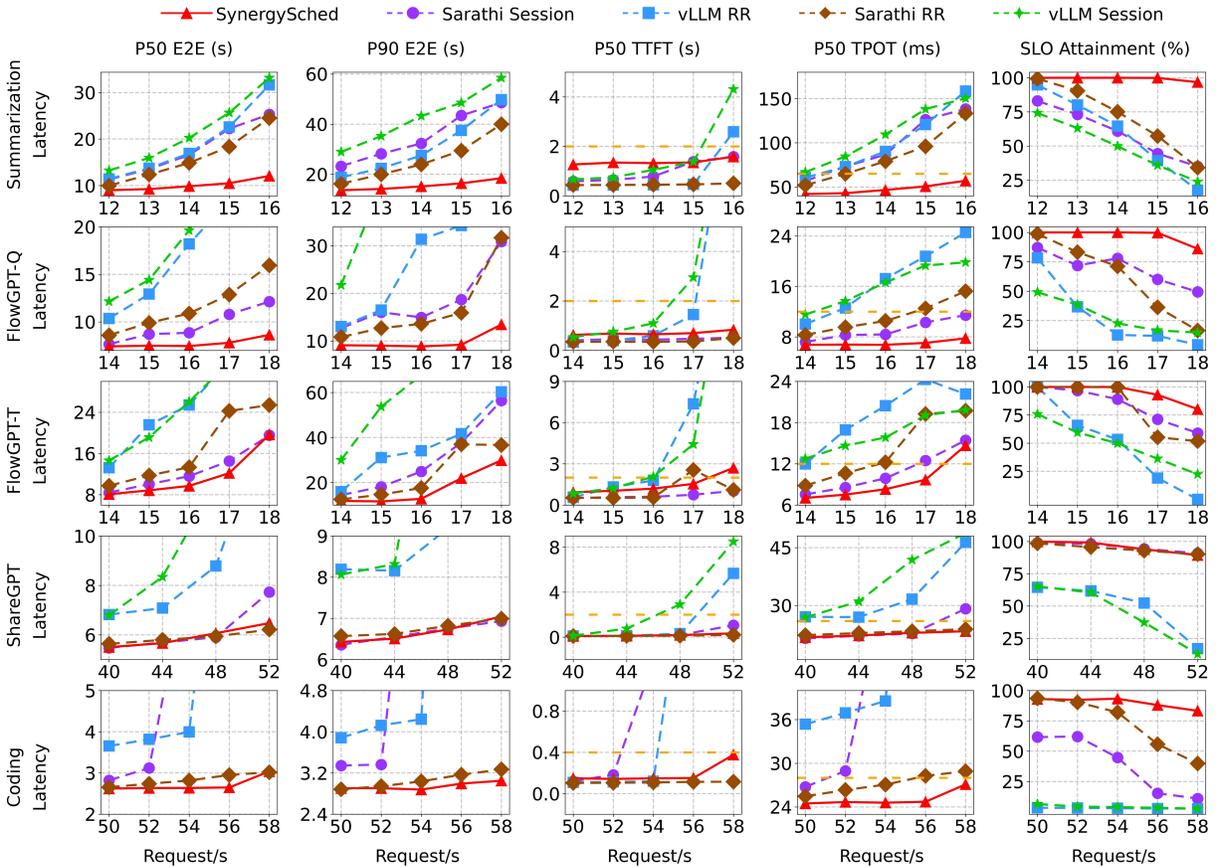}
\caption{End-to-End evaluation. (The yellow dashed line in the TTFT and TPOT plots indicates the specified SLO for that workload.)}
\label{fig:e2e}
\end{figure*}

We evaluated NexusSched's end-to-end performance on a homogeneous 8$\times$H100 cluster to demonstrate that bridging the two information gaps directly translates to superior performance and stability. As shown in Fig.~\ref{fig:e2e}, our predictive, cross-layer co-design significantly outperforms all baseline systems across every metric.


Fig.~\ref{fig:e2e} shows that for all five scenarios with different characteristics, NexusSched is almost always at the lowest level for all latency metrics, with a flatter growth curve. This translates to a 20\%--50\% reduction in P50 and P90 end-to-end latency and an average SLO attainment improvement of over 43\%. The impact is most dramatic under high load; for instance, in the compute-intensive Summarization scenario, NexusSched serves over 3$\times$ as many requests as the best baseline under SLO constraints. This advantage is a direct result of our synergistic design: LENS's SLO-aware scheduling bridges the internal gap by adapting to the heavy loads, while PRISM's predictive routing bridges the external gap by making informed routing. In contrast, the baselines' performance collapses because their reactive, uninformed strategies cannot cope with dynamic, high-load traffic.

\subsection{Ablation Study}
\label{sec:ablation}

To quantify the performance impact of bridging each information gap, we conducted an ablation study on the 8$\times$H100 cluster. We compared four configurations: \textbf{Sarathi+RR} where both gaps are open; \textbf{LENS+RR}, which closes only the internal gap; \textbf{Sarathi+PRISM}, which attempts to close the external gap with limited, non-predictive state signals; and the full \textbf{NexusSched} system, which closes both gaps synergistically.

\begin{figure}[!t]
\centering
\includegraphics[width=0.8\columnwidth]{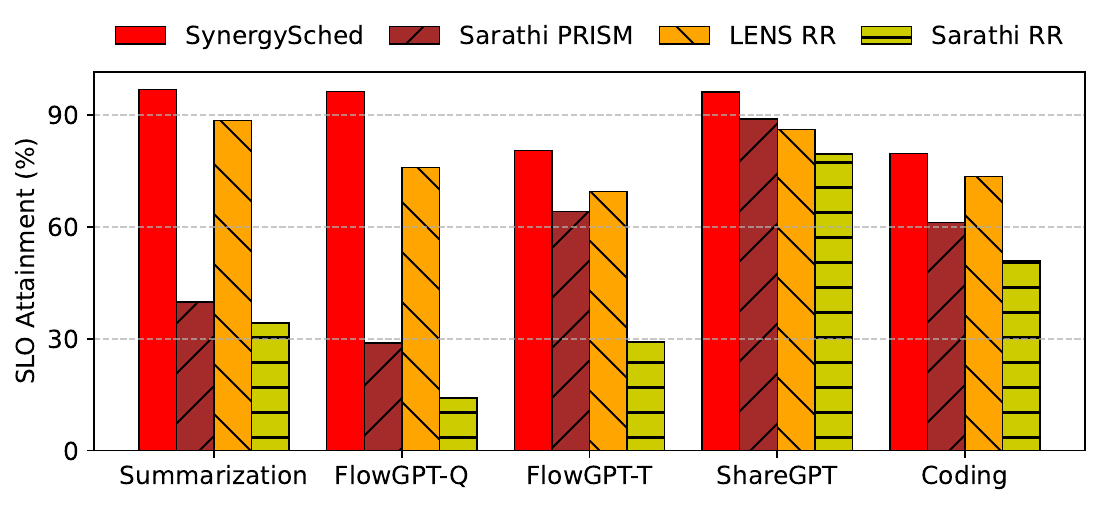}
\caption{Ablation study results.}
\label{fig:ablation}
\end{figure}

We conducted this experiment under a high load, slightly below NexusSched's saturation point. The results, shown in Fig.~\ref{fig:ablation}, quantitatively prove our core thesis. We found that a standalone LENS and an information-limited PRISM improve the SLO attainment rate by an average of 2.5x and 1.4x respectively. However, the impact of different components varies significantly across workloads. LENS's gains are most prominent in compute-intensive workloads like Summarization, delivering up to a 3.1x improvement. PRISM’s standalone gains are modest, with its main advantage seen in scenarios like FlowGPT-T that require balancing session affinity and load under bursty traffic. This confirms that even an intelligent router is ineffective without high-quality, predictive states from the engine. The complete NexusSched system achieves the highest performance by a significant margin, demonstrating that maximum efficiency is only unlocked when \textbf{both information gaps are bridged synergistically}.

\subsection{Heterogeneous Cluster Performance}
\label{sec:heterogeneous_cluster_performance}

\begin{figure*}[!t]
\centering
\includegraphics[width=2\columnwidth]{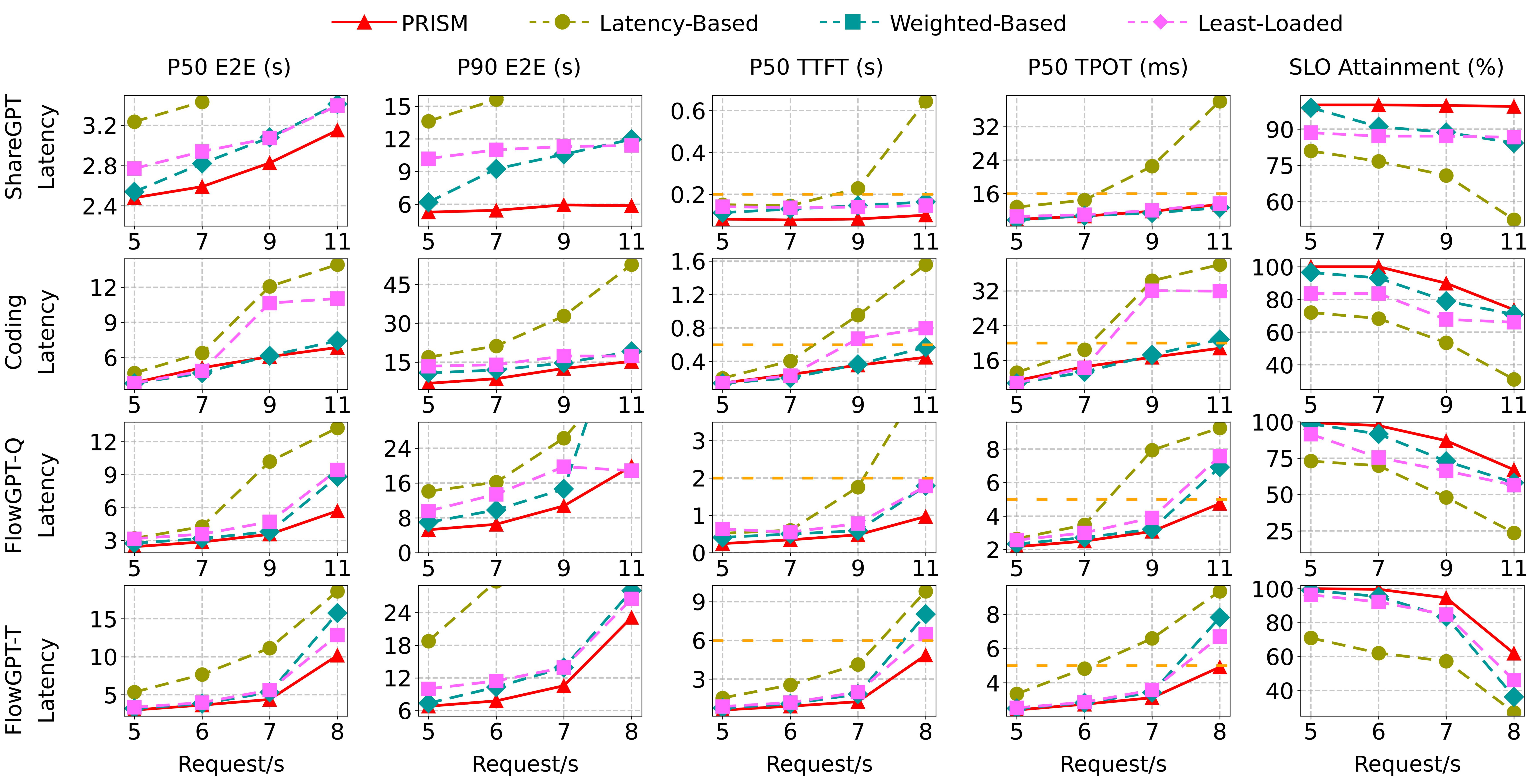}
\caption{Performance on a heterogeneous cluster. (The yellow dashed line in the TTFT and TPOT plots indicates the specified SLO for that workload.)}
\label{fig:heterogeneous}
\end{figure*}

This experiment evaluates NexusSched's core advantage in heterogeneous clusters: its ability to perceive each instance's service capability and combine it with its dynamic, real-time load conditions to make globally optimal routing decisions. To provide a comprehensive comparison, we evaluate NexusSched against three representative routing policies common in industry, all paired with our underlying LENS engine: a dynamic \textbf{Latency-Based} policy using recent 2\,s average latency; a static \textbf{Weighted-Based} policy using offline-profiled QPS weights; and a \textbf{Least-Loaded} policy using queue length as a proxy for load. Due to the specific hardware constraints of the heterogeneous cluster, this experiment uses the Llama3-8B model and excludes the resource-intensive Summarization scenario.

The results in Fig.~\ref{fig:hete_sched_overview} and Fig.~\ref{fig:heterogeneous} strongly demonstrate the superiority of this approach. NexusSched exhibits the best performance and stability, reducing P50 end-to-end latency by 7\% to 36\% and maintaining 10\%-25\% higher SLO attainment at high load points. Fig.~\ref{fig:qps_share_flowgpt_ts} reveals the fundamental inefficiency of the other policies: Weighted-Based is a purely static policy that completely ignores real-time load, resulting in suboptimal performance. Latency-Based and Least-Loaded are typical reactive policies. They not only lack an understanding of node capabilities but also depend on lagging metrics, causing them to frequently make poor decisions under bursty scenarios. In contrast, PRISM uses the rich, forward-looking state exported by each engine. This allows it to precisely match requests to the true, near-future availability of each node, thereby maximum effective throughput of the entire cluster.

\subsection{Zero-Config Deployment}
\label{sec: zero-config}

\begin{figure}[!t]
\centering
\includegraphics[width=0.7\columnwidth]{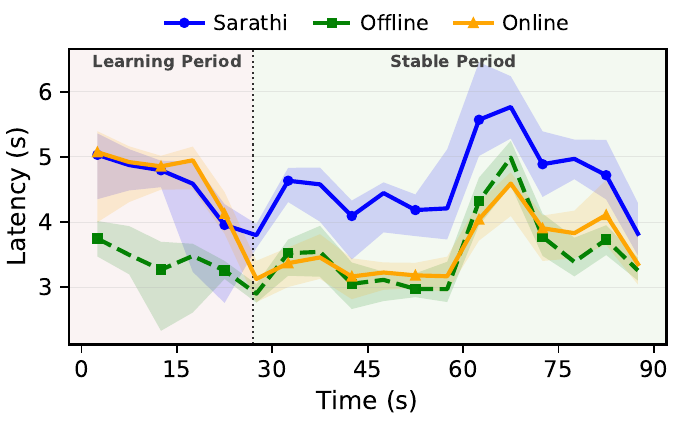}
\caption{Performance in zero-config deployment.}
\label{fig:online_learning}
\end{figure}


This experiment validates a key feature of NexusSched: its ability to ``plug-and-play'' with new, unprofiled resources by rapidly adapting through online learning. We simulate this scenario by introducing an ``unknown'' H100 GPU into a running cluster and immediately placing it under a high-intensity request stream. We compare NexusSched with online learning against a baseline Sarathi system and a theoretically optimal, offline-profiled NexusSched instance.

As shown in Fig.~\ref{fig:online_learning}, the results are definitive. While initially performing poorly due to an inaccurate performance model, NexusSched's online optimizer quickly self-corrects by continuously learning from runtime performance data and adjusting its model parameters in real time. Consequently, its average request latency drops sharply and converges to the ideal, offline-optimized performance in 30 seconds. This rapid self-adaptation capability confirms that NexusSched can efficiently utilize transient, heterogeneous resources like spot instances, significantly improving the cost-effectiveness and operational agility of large-scale LLM serving.

\subsection{Micro-benchmark Analysis}
\label{sec: micro}

\begin{figure}[!t]
\centering
\begin{subfigure}[t]{0.48\columnwidth}
    \centering
    \includegraphics[width=\linewidth]{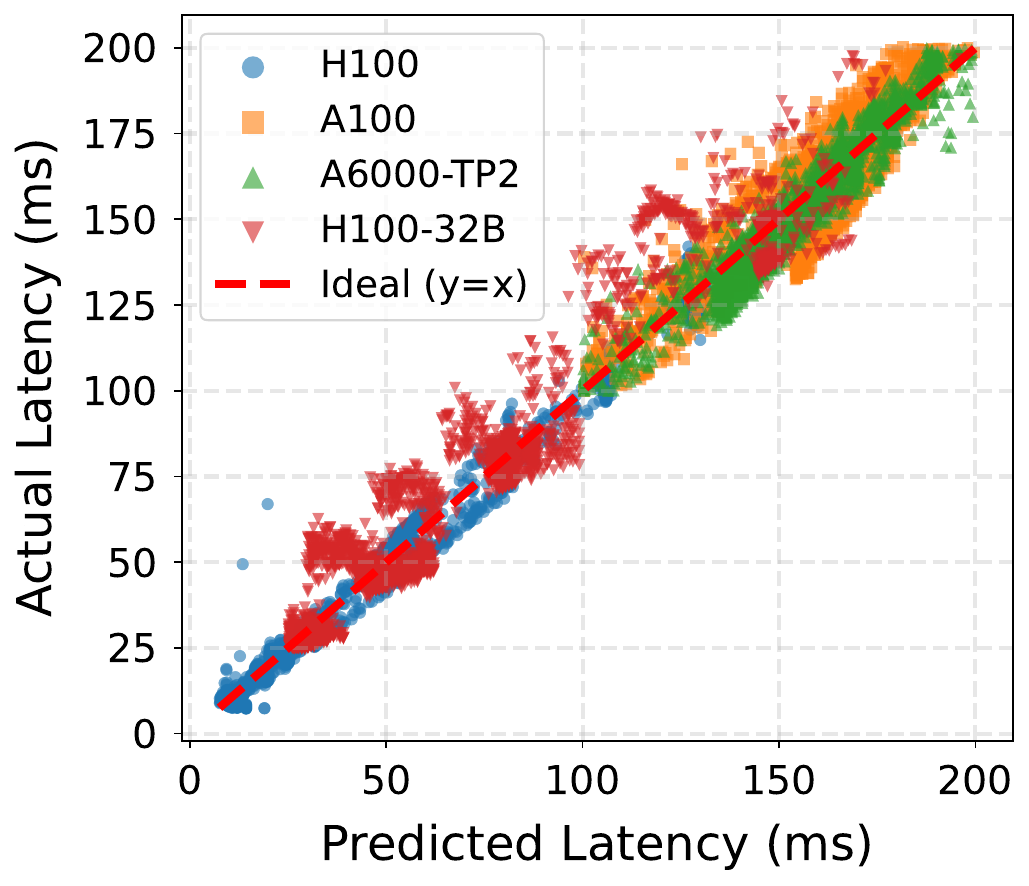}
    \caption{Predicted Latency vs. Actual Latency (Llama3-8B if unmarked).}
    \label{fig:predicted_and_actual_latency}
\end{subfigure}
\hfill
\begin{subfigure}[t]{0.48\columnwidth}
    \centering
    \includegraphics[width=\linewidth]{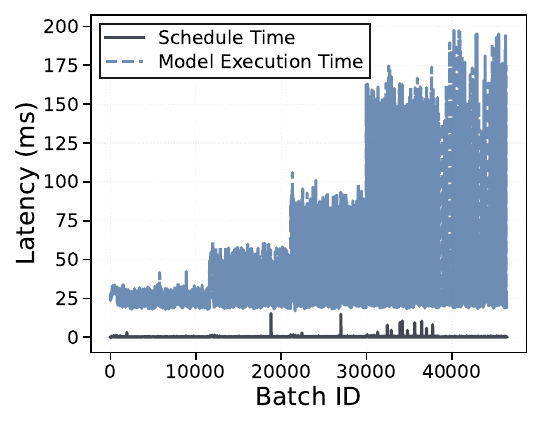}
    \caption{Scheduling Overhead vs. Model Execution Time per iteration.}
    \label{fig:scheduling_overhead}
\end{subfigure}
\caption{(a) The performance model demonstrates high predictive accuracy; (b) NexusSched's scheduling overhead proved to be negligible.}
\label{fig:model_accuracy_overhead}
\end{figure}


\subsubsection{Performance Model Accuracy}

The effectiveness of our proactive scheduling framework rests on the accuracy of our performance model. To validate its predictive power, we compared its predictions against thousands of profiled latencies from batches with varying configurations across heterogeneous GPUs (H100, A100, A6000) and models (Llama3-8B and QwQ-32B). As shown in Fig.~\ref{fig:predicted_and_actual_latency}, the results demonstrate an exceptionally strong correlation. The data points cluster tightly along the ideal y=x diagonal, achieving a coefficient of determination ($R^2$) between 0.95 and 0.99. This high fidelity provides a reliable foundation for all of NexusSched's scheduling and routing decisions.



\subsubsection{Scheduling Overhead}

Last but not least, a critical design goal is that our methods should not become a new performance bottleneck. Our analysis confirms that NexusSched's overhead is negligible. As visualized in Fig.~\ref{fig:scheduling_overhead}, the engine-layer LENS scheduling decisions consistently take under 1\,ms, which is orders of magnitude lower than the GPU computation latency (20--200+\,ms). Separately, the cluster-layer PRISM routing algorithm adds approximately 1\,ms of overhead per request. Together, these results confirm that our framework's decision-making logic is lightweight and imposes no significant overhead on the critical serving path.

\section{Related Works}

\textbf{LLM Serving Engines.}
LLM engines have evolved from request-level batching to iteration-level, objective-driven scheduling.
Request-level batching~\cite{nvidia2021fastertransformer} improves theoretical throughput but induces head-of-line blocking.
Orca~\cite{yu2022orca} introduces iteration-level scheduling and selective batching.
vLLM~\cite{kwon2023efficient} couples PagedAttention with prefill-priority to raise occupancy and reduce TTFT.
To further lift throughput, Sarathi-Serve~\cite{narayanan2024sarathi} and DeepSpeed-FastGen~\cite{holmes2024deepspeed} employ stall-free chunked prefill to reduce head-of-line blocking.
More recent work has explored architectural changes like disaggregating prefill and decode stages~\cite{zhong2024distserve} or incorporating explicit SLO-awareness using simple cost models~\cite{chen2025slosserve,bin2025fineserve,hong2025sola}.
Our work differs by placing a structure-aware, online performance model at the core of iteration-level decisions.

\textbf{Cluster-Layer Request Routing.}
Routing frameworks operate at the macro level across many engines.
Homogeneous clusters emphasize fairness/utilization, such as Ray~\cite{moritz2018ray} and Production-Stack~\cite{productionstack2025}, while heterogeneous settings require capability-aware routing, such as AIBrix~\cite{xu2024aibrix}, LLM-D~\cite{llmd2025}, and DynamoLLM~\cite{stojkovic2024dynamollm}.
However, these systems are fundamentally limited by an information gap, as they rely on coarse-grained or lagging metrics (e.g., queue length, past average latency) which fail to reflect an engine's true, near-future capacity. In contrast, NexusSched closes this gap by introducing a predictive approach, using the forward-looking state exported from each engine to enable proactive, performance-aware request orchestration.

\textbf{Performance Modeling and Predictive Scheduling.}
Prediction has guided cloud autoscaling and resource management (e.g., SageServe~\cite{chakrabarti2025sageserve}, Aladdin~\cite{zhu2024aladdin}, SkyLB~\cite{skylb2025}), but these approaches target macro elasticity and lack batch-level fidelity and hardware generality.
Within LLM inference, polynomial regressions~\cite{chen2025slosserve,hong2025sola,goel2025niyamabreakingsilos} and operator-centric models (FlashAttention~\cite{dao2023flashattention}, Optimum/DeepSpeed-Inference~\cite{optimum,deepspeed-inference}) provide useful estimates yet are not used into SLO-aware iteration scheduling or capacity-aware routing.
NexusSched addresses this gap with a performance model that captures saturation and token-dependent effects, driving both intra-engine scheduling and inter-engine routing in LLM serving.

\section{Conclusion}

This paper identifies two fundamental information gaps in LLM serving and argues for a paradigm shift from reactive scheduling to a predictive, cross-layer co-design to solve them. We presented \textbf{NexusSched}, a framework built on a structurally-informed online performance model. Its engine-layer scheduler, \textbf{LENS}, bridges the internal gap with SLO-aware adaptive batching, while its cluster-layer router, \textbf{PRISM}, bridges the external gap with predictive orchestration. Our evaluations show that this approach reduces end-to-end latency by up to 50\% and improves SLO attainment by over 43\%. These results confirm that a synergistic, predictive co-design is not merely beneficial, but essential for building robust and scalable LLM serving systems.

\bibliographystyle{ACM-Reference-Format}
\bibliography{sample-base}

\appendix

\end{document}